\documentclass[11pt]{article}
\usepackage{moriond,epsfig}

\def\Journal#1#2#3#4{{#1} {\bf #2}, #3 (#4)}

\def\PLB{{\em Phys. Lett.}  B}
\def\PRL{\em Phys. Rev. Lett.}
\def\PRD{{\em Phys. Rev.} D}

\def\be{\begin{equation}}
\def\ee{\end{equation}}
\def\bea{\begin{eqnarray}}
\def\eea{\end{eqnarray}}

\begin{document}
\vspace*{4cm}
\title{TOP MASS AND PROPERTIES}

\author{ Yen-Chu Chen \\ On behalf of CDF and D0 collaboration}

\address{Institute of Physics, Academia Sinica, \\
Taipei, Taiwan, Republic of China}

\maketitle\abstracts{
The top quark was discovered in 1995.  The top quark mass is now well
measured at the Tevatron, with uncertainty getting below 1\% of the top
mass.  The world average from last year was 170.9 $\pm$ 1.8 GeV/$c^2$.
The new CDF measurement is 172 $\pm$ 1.2 (stat) $\pm$ 1.5 (sys) GeV/$c^2$,
and D0 will soon present a new measurement.
The top quark mass is an important parameter in the Standard Model,
and should be measured as precisely as possible.
To learn more about the top quark observed and study possible
new physics, other properties also should be measured.
At the Tevatron, the charge of the top quark can be measured directly. 
Examples of other properties studied and reported in this presentation
are W helicity, top decay branching ratio to b ($R_b$), searches for
$t \rightarrow H b$ and for flavor changing neutral current (FCNC).
The results are all consistent with the Standard Model
within current statistics.  
With significantly more data being collected at the Tevatron,
precision measurements of the top properties are just starting.
}

\section{Introduction}

Top quarks are produced at the Tevatron
mainly in top anti-top pairs,  $P\bar{P} \rightarrow t\bar{t}$,
through quark anti-quark annihilation and gluon gluon fusion.
The $t$ $(\bar{t})$ quark subsequently decays into 
  a $W^+(W^-)$ boson and a $b(\bar{b})$ quark,
    $t \rightarrow W b $, with a branching ratio close to 1. 
From the $b$'s and the final products of the $W$ decays,
the mass and other properties of the top quark can be measured. 

The $P\bar{P} \rightarrow t\bar{t} \rightarrow W^+b W^-\bar{b}$
production cross section and event selection
have been reported in a previous talk at this Conference. 
Based on how the $W$'s decay, three analysis channels are identified:
  the di-lepton channel (DIL) for both W's decaying leptonically,
  the lepton plus jet channel (LJ) for only one W decaying leptonically,
  and the all hadron channel for both $W$'s decaying hadronically.
(In this article we consider the final leptons being electron or muon 
 only.  The case of $W \rightarrow \tau \nu $
 has to be handled differently due to the nature of tau decay.)
Each channel has its own challenges and strengths.
Some common methods are developed and used when applicable. 

\section{top mass measurement}

Mass is a fundamental property of a particle. 
While the top has been discovered for more than ten years,
we have many interesting questions about the top quark.
Are we seeing the same particle in all three analysis channels
 (DIL, LJ and all hardon)? 
Precise measurement of the top mass in these three channels
could provide some insight to this question. 
If this particle seen as top quark is truly the one of the SM, then
since its mass strongly correlates with the mass of Higgs particle, 
precise measurement of the top mass can help the search of the
Higgs particle, and will also enable stringent constraints for
electroweak tests and new physics.

The three channels, DIL, LJ and all hadron, have their own challenges
and methods.  In the DIL channel, each event has two neutrinos
that are not measured directly, with only the missing transverse
energy providing partial information for these two neutrinos.
The system is under-constrained.
Additional assumption is used to further constrain
the system to be able to reconstruct the top mass.  Also, 
various top mass could be used as input to obtain a probability
density function to determine the most probable value for top mass.

A general issue with all three channels is: 
which lepton and jet(s) in each $t\bar{t}$ event are decay
products of the top quark and which are from the anti-top? 
One could try all possible combinations and select one based
on reconstruction probability or simple kinematic information,
such as the invariant mass of the top and anti-top system. 
Alternatively, one could use all possible solutions and assign 
weights based on the some relevant quantities, such as a weight
defined by comparing missing Et from the reconstruction to that
from what is measured in each event.  These techniques are
also generally applied in studying properties of the top quark.


\subsection{The methods}\label{subsec:insitu}

One common issue with all channels is the jet energy calibration.
In prior analysis jet energy calibration was based on predefined
cone sizes.  In an event where at least one $W$ decays hadronically,
the known $W$ boson mass can be used as input to further fine tune
the two jets associated with this $W$.
This is called the in-situ jet energy calibration.
In top mass measurement, this W mass constraint is applied to the
final events selected to find the average shift to the nominal jet energy 
calibration.  The shift is applied to all jets including the $b$ jets.
This procedure significantly improves the determination of the uncertainty 
in the top mass measurement.

The Template Method is one of the main methods used to obtain the top 
mass.  In this method, top mass is reconstructed from the kinematic 
information available in the event.  Templates are formed based on
Monte Carlo with different top mass input.  Comparing these templates
with the observed events reconstructed in the same way reveals the
top mass.  In each DIL channel event, the system is under-constrained. 
Additional reasonable assumption has to be made, such as taking 
$P_z$ of top anti-top system from observed events~\cite{p:pztt}, or 
weighting on the $\phi$ angle of the neutrino~\cite{p:phiweight}, etc.
In the LJ channel, the assumption is made that the missing energy
is due to the neutrino being not detected.  A top mass fitter is 
used to find the most probable top mass, taking into account the 
resolution of $p_t$ and jet energy measured.  In-situ jet energy
calibration is commonly applied to improve the uncertainty. 
In each all hadronic event, there are two $W$'s decaying hadronically.
In-situ jet energy calibration is generally applied to the jets
which form the $W$'s.  In the Template Method with 2-dimensional fit 
(TMT2D)~\cite{p:cdftmt2d} analysis in CDF, all possible jet pairing 
combinations are tried but only the one with best $\chi2$ is kept.

The Matrix Element Method is based on theory.  This takes 
into account all the kinematics information contained in an event, 
which are top mass dependent. A conditional probability can be formed 
for a given top mass. In DIL, this probability can be expressed as 

\begin{equation}
P(\mathbf{x}|M_t) = \frac{1}{N}\int d\Phi_8 |{\mathcal M}_{t 
\overline{t}}(p;M_t)|^2  \prod_{jets} W(p,j) f_{PDF}(q_1)f_{PDF}(q_2), 
\label{eq:meprob}
\end{equation}

\noindent where $M_t$ is the top mass, 
$\mathbf{x}$ contains the lepton and jet energy measurement, 
${\mathcal M}_{t \overline{t}}(p;M_t)$ is the
    $t\bar{t} $ production matrix~\cite{p:mematrix}, 
$q$ is the vector of incoming parton-level quantities, 
$p$ is the vector of resulting parton-level quantities: 
    lepton and quark momenta, 
$W(p,j)$ is the transfer function which gives the probability
    to observe a jet with energy $j$ given a parton energy $p$, 
and finally, $f_{PDF}$ the parton distribution functions of
    the two quarks from the proton and anti-proton.
The integral is over the entire six-particle phase space. 
Scanning through the top mass,
 the most probable point reveals the mass of the top quark. 
An example of applying such method for top mass
measurement is performed at CDF using DIL samples~\cite{p:cdfdilme}.

This method ``Matrix Weighting" is different from the 
            ``Matrix Element" method described previously. 
This method is applied to DIL samples, where the system is 
under-constrained due to missing neutrinos. 
For a given top mass, one could try to resolve for $t \bar{t}$ momentum.
A weight is calculated for each solution found by comparing the missing 
energy calculated with the one observed in observed events.
The top mass is determined from a scan through a range of top mass 
to find a maximum weight and the extremum of likelihood. 
This is described in D0's public conference note~\cite{p:d0mtw}. 

``Neutrino weighting" is a method applied in D0. 
Using DIL samples, for a given top mass $\eta$ 
was thrown based on Monte Carlo simulation for each $\nu$. 
Then the set of energy-momentum conservation equations 
can be resolved for $\nu$ momentums. 
For each event a weight template was derived based on missing energy 
expected and observed at each given top mass. 
A maximum likelihood is formed, combining all events, and 
the extremum of this distribution reveals the top mass. 
This is described in D0's public conference note~\cite{p:d0nuw}.

At the Tevatron, many techniques have been developed to measure the
top mass.  Progress has been made to improve the uncertainty.
Some of the methods have not been mentioned in this presentation.
A single variable that has a distribution being sensitive to the
top mass can be used to do the measurement. One such variable is
the Lxy, which is the closet distance of the secondary vertex to 
the primary vertex in the transverse plan of the detector.
may have a distribution which is sensitive to the top mass.
The top mass measurement from the top production cross section 
is discussed by Marc Besancon at this Conference.
All of the methods provide additional info, and
could help in improving uncertainty of the combined top mass.

\subsection{The results}

\begin{table}[t]
\caption{top mass measurement at the Tevatron \label{tab:tmass}}
\vspace{0.4cm}
\begin{center}
\begin{tabular}{|c|c|c|}
\hline
& & \\
Analysis&
Samples&
Result 
\\ \hline
ME+NN (CDF)~\cite{p:cdfdilme} &
DIL, 2 fb$^{-1}$ &
171.2 $\pm$ 2.7($stat$) $\pm$ 2.9($sys$) GeV/$c^2$ 
\\ \hline
TMP+NN (CDF)~\cite{p:cdftmt2d} & 
Had., 1.9 fb$^{-1}$ &
177.0 $\pm$ 3.7($stat+JES$) $\pm$ 1.6($sys$) GeV/$c^2$ 
\\ \hline
ME+NN (CDF)~\cite{p:cdfmelj} & 
LJ, 1.9 fb$^{-1}$ &
172.7 $\pm$ 1.2($stat$) $\pm$ 1.3($JES$) $\pm$ 1.2($sys$) GeV/$c^2$ 
\\ \hline
MW (D0)~\cite{p:d0mtw} &
DIL, 1 fb$^{-1}$ &
175.2 $\pm$ 6.1($stat$) $\pm$ 3.4($sys$) GeV/$c^2$
\\ \hline
NW (D0)`\cite{p:d0nuw} &
DIL, 1 fb$^{-1}$ &
172.5 $\pm$ 5.8($stat$) $\pm$ 3.5($sys$) GeV/$c^2$
\\ \hline
\end{tabular}
\end{center}
\end{table}

The results on the top mass measurement at Tevatron given at this 
conference are listed in Table \ref{tab:tmass}. 
All individual top mass measurement from all three channels
show consistent results. 
There is no indication of seeing different particles in different 
channels.

\begin{figure}
\psfig{figure=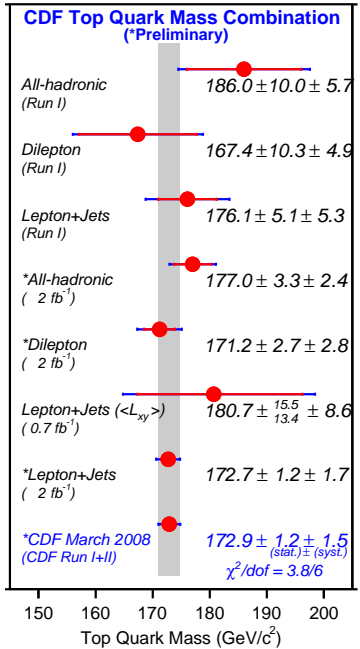,height=3.5in}
\hskip 1.5 cm
\psfig{figure=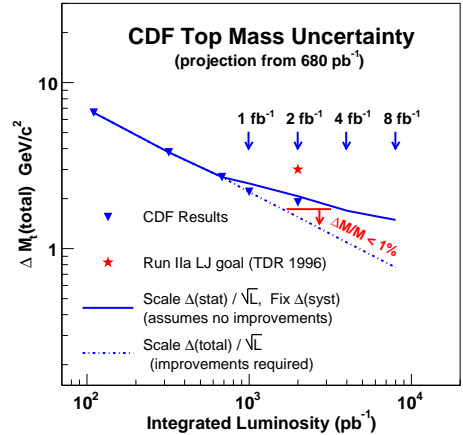,height=2.5in}
\caption{The combined measurement of top mass from CDF. 
The plot on the right shows the improvement of uncertainty with
respect to the integrated luminosity. 
The dark blue points are the reality, compared to the projection
based on including more data only (blue line) or further 
improving the analysis methods (dashed line). 
The improvement based better methods is hard to predict. 
The dashed line is the most optimistic case.
The new results are between the two lines.
CDF alone is at the same level as CDF and D0 combined last year. 
Combining effort from CDF and D0, the uncertainty of top mass
should be less than 1\% of the measured top mass. 
\label{fig:tmasscdf}}
\end{figure}

At the moment of this presentation, CDF has already a combined 
result using various results from all hadronic, DIL and LJ channels.
This yields  $172.9 \pm 1.2 (stat) \pm 1.5(sys)~ \rm{GeV/}c^2$
 and is shown in Figure \ref{fig:tmasscdf}. 
This result is approaching an uncertainty of 1\% of the top mass, 
which is similar to CDF and D0 combined result for the year 2007.
 Together with the updated D0 measurement, the new 
 combined result would have an uncertainty below 1\%.
(This happened right after the 
   Moriond EW 2008 conference.~\cite{p:TopMassCDFD02008March}.)
CDF and D0 are working together on common systematic issues
to improve uncertainty at the Tevatron for the high precision era of
   top mass measurement.

\section{Top property studies}
The SM top quark has spin ($1/2$), charge ($+2/3$),
and other definite properties which should be measured.
In contrast, the top mass is a free parameter in SM.
Any significant deviation of the top quark properties from SM
would indicate new physics.
The top charge is among the fundamental properties of the
top quark most accessible at Tevatron. 
Other properties, such as top spin, lifetime, decay width,
  either need significantly more data
  or are far beyond our capability to measure with our given 
detector resolution.
Studies from the top deccay
 include the helicity of W boson from top decay, 
 measurement of branching ratio, search for charged Higgs, 
 search for flavor changing neutral current, etc. 

\subsection{The charge of top quark}
In the SM, the charge of top quark is $+2/3$. 
An alternative possibility
suggested by an exotic model (XM)~\cite{p:xmchang} is $-4/3$. 
In this model, it is claimed that the particle seen
at Tevatron may be an exotic top of charge
$-4/3$, which decays into $W^-$ and $b$, 
unlike in the SM where the top quark decays into $W^+$ and $b$. 
The two key elements in the study are then to identify the
source of a jet being $b$ or $\bar{b}$, and 
how the $b$ and $\bar{b}$ jets are paired with the two $W$'s.
 
The identification of a jet being from $b$ or $\bar{b}$ is
done via calculating the jet charge, which is sum of
jet-track charges weighted by the track momentum amplitude 
and how close the track is to the jet axis. 
For true $b$ jets this method has 60\% probability of 
identifying $b$ or $\bar{b}$ correctly. 

The pairing can be done by taking the measured top mass as input
and check which pairing is more probable. In DIL channel,
events can be selected based on the square of invariant mass 
of the paired lepton and jet $m_{lb}2$ to improve the purity.
In each event there are two possible ways of pairing and four 
possible $m_{lb}2$ values. The pairing having the maximum 
$m_{lb}2$ does not always provide the correct pairing. 
In the events with the maximum $m_{lb}2$ is greater
than certain value, this method can be almost 100\% correct. 
However cutting too tight would lose too much in statistics.
The best point for making such cut is 21000, 
assuming that SM is true and top mass is 175 GeV/$c^2$. 
With this selection, 94\% of pairing purity can be reached
  with efficiency of 39\%.

The charge of the top quark was first studied by D0. 
With 0.37 fb$^{-1}$ of data, the result prefers the SM 
instead of XM~\cite{p:d0topq}. 
In CDF, the study has been done with data up to 1.5 fb$^{-1}$. 
The result~\cite{p:cdftopq} up to date support SM over XM, 
and the XM is rejected at 87\% confidence level (CL). 
Combining DIL and LJ, among 225 top or anti-top quark decays
 124 decays support SM and 101 support XM. 
Correcting for purity of the analysis, the measured true 
fraction of SM over total is 0.87, 
which based on our sensitivity gives a p value of 0.31. 
An additional way of showing this is the Bayes Factor $(BF)$,
which is defined as $P(N_+|SM)/P(N_+|XM)$, 
i.e. the probability of observed events 
happening assuming SM is true over the one of XM. 
A common way to utilize $BF$ is the quantity $L = 2*Ln(BF)$. 
For $L$ in the ranges (0-2), (2-6), (6-10), ($>$10),   
    the result is uncertain, positive, strongly supporting SM,
                or very strongly supporting SM, respectively.
Our result from CDF data is 12.01,
    thus very strongly support SM over XM.
With more data, we will determine more precisely the top charge.

\subsection{W helicity}
In the SM, V-A rules the weak decay. 
The $W$ boson from top quark decay is thus polarized. 
The SM predicts that the $W$ helicity
in this case should have 70\% 
longitudinal ($f_0$) and 30\% left-handed ($f_-$). 
The component of right-handed 
($f_+$) is very small, 3.6 $\times$ 10$^{-4}$. 
Significant deviation of $f_+$ would indicate new physics.

The study of W helicity can be performed via looking at the
 $cos \theta^*$ distribution,
where $\theta^*$ is the angle of the electron or muon in the
 $W$ rest frame with respect to the anti-direction of top quark
 in this frame. The analysis can be performed in LJ and DIL channels. 
In case of LJ the missing energy is assumed to be due to the missing 
neutrino. Events can be reconstructed using top mass as
input and lepton angle in W rest frame can be calculated. 
The top mass used is generally 175 GeV/$c^2$. 
In case of DIL there are two missing neutrinos. 
Using top mass as input one can figure out which jet is paired
with which lepton and resolve for the neutrino momenta. 
Lepton angle in the W rest frame can be obtained in this way. 
CDF does this analysis, using 1.9 fb$^{-1}$ data, 
in the LJ channel. 
In a 2 dimensional fit where both $f_0$ and $f_+$ are 
fitted at the same time 
the result shows $f_0 = 0.65 \pm 0.19 (stat) \pm 0.03 (sys)$ 
and $f_+ = -0.03 \pm 0.07 (stat) \pm 0.03 (sys)$. 
If $f_0$ is fixed to the SM value CDF obtains 
$f_+ = -0.04 \pm 0.04 (stat) \pm 0.03 (sys) $ 
and sets upper limit for $f_+$ at 0.07 at 95\% CL~\cite{p:cdfwhel}. 
D0 collaboration does the analysis in both LJ and DIL channels. 
A 2-D fit of $f_0$ and $f_+$ reveals 
 $0.425 \pm 0.166 (stat) \pm 0.102 (sys)$ 
and $0.119 \pm 0.0090 (stat) \pm 0.053 (sys)$ respectively. 
Fixing $f_0$ to the SM value
gives $f_+ = -0.002 \pm 0.047 (stat) \pm 0.047 (sys)$. 
An upper limit of 0.13 at 95\% CL is set~\cite{p:d0whel}.

\subsection{Study of $R_b$}
A study on the 
     $R_b = Br( t \rightarrow W b ) / Br( t \rightarrow W q )$, 
where $q$ represents all possible quarks allowed in the decay, 
is performed at D0. 
$R_b$ is correlated with the top pair production. 
Noting that D0 does simultaneous fit to both values, 
using LJ channel from 0.9 fb$^{-1}$ data~\cite{p:d0rb}. 
The result is $R_b = 0.97^{+0.09} _{-0.08} (stat+sys)$. 
A lower limit of $R_b$ is set at 0.79 at 95\% CL.
From this a lower limit on $|V_{tb}|$ is set at 0.89 at 95\% CL.
From the same fit the resulted production cross section is 
$\sigma_{t \bar{t}} = 
    8.18 ^{+0.90}_{-0.84} (stat+sys) \pm 0.50 (lumi)$ pb,
 which is consistent with the direct measurement.

\subsection{Search for $t \rightarrow H b$ }
It is interesting to search for charged Higgs in the top quark
decay.  D0 collaboration did this analysis by comparing the
 production cross section of top pair from the LJ channel
against the one from the DIL channel. 
If there were charged Higgs in the top decay, it would mostly 
contribute to the LJ channel but much less in the DIL channel. 
The ratio of the two production cross sections 
 is  $R = 1.21 ^{+0.27} _{-0.26} (stat+sys)$, 
based on the assumption that $R_b = 1$. 
Extracting the branching ratio of $t \rightarrow H b$ from this
cross section ratio, 
D0 obtains $Br = 0.13 ^{+0.12} _{-0.11} (stat+sys)$. 
An upper limit is set at 0.35 at 95\% CL~\cite{p:d0thb}.

\subsection{Search for FCNC}

At CDF an analysis to study FCNC is to 
   search for $t \rightarrow Z q$ in the top quark decay. 
The SM predicts a branching ratio at the order of $O(10^{-14})$. 
However beyond SM up to $O(10^{-4})$ is possible. 
At CDF events having two high $p_t$ leptons with at least 
4 jets were selected with constraint on masses of top, Z and W.
Comparing the data (1.9 fb$^{-1}$) with expectation,
   no excess is seen. 
An upper limit is set at 3.7\% at 95\% CL~\cite{p:cdffcnc}.

\section{Summary and Future Prospects}

The top quark mass has been well measured at the Tevatron,
with uncertainty getting below 1\% of the top mass.  
The top quark mass is an important parameter in the Standard Model,
and should be measured as precisely as possible.
Other properties of the top quark also should be measured,
to learn more about the top quark and study possible new physics. 
Examples of other top studies at the Tevatron are 
   the charge of the top quark,
   W helicity, top decay branching ratio to b ($R_b$), searches for
   $t \rightarrow H b$ and for flavor changing neutral current (FCNC).
The results are all consistent with the Standard Model
within current statistics.
With significantly more data being collected at the Tevatron,
precision measurements of the top properties are just starting.

\section*{Acknowledgments}
Many thanks to the people working on Tevatron. With their dedicated
effort to improve the accelerator discoveries are made possible.
Many thanks to the funding agencies to CDF and D0 collaboration. 
Many thanks to the authors and conveners of both collaboration who provided 
critical input to this presentation. At the end I want to thank the 
organizers of the Moriond EW 2008 conference for their warm hospitality 
and well organized program.

\end{document}